\documentstyle[aps,prl,multicol,epsf]{revtex}

\begin{document}
\draft
\preprint{}

\title{The superconductor insulator transition in systems of ultrasmall grains}
\author{A. Frydman}
\address{The Minerva Center, The Physics Department, Bar Ilan University, Ramat Gan\\
52900, Israel}
\date{\today}
\maketitle

\begin{abstract}
We present transport measurements on quench condensed granular Pb films in
which the grains are $40-80$\AA\ in diameter. These films show a cross
over from an insulator to a superconductor behavior as the nominal thickness
of the layer is increased. This transition is different in nature than those
seen in quench condensed systems reported in the past where the films were
either uniform or granular with grain sizes on the order of $200$\AA . We
discuss possible physical mechanisms for these transitions.

\end{abstract}

\begin{multicols}{2}
The superconductor insulator transition (SIT) in thin layers has attracted a
lot of interest for close to three decades and has been lately revived due
to the possibility that it is a quantum phase transition at T=0 \cite
{hebard,kapitulnik,gantmacher,fisher,fermion,goldman today}. A number of
theoretical scenarios have been suggested to try to explain why
superconductivity is destroyed as the thickness of the layer is decreased.
These can be classified into two major classes. The first invokes a
''Bosonic'' picture in which isolated superconducting islands exist in a non
superconducting matrix\cite{fisher}. Here, Cooper pairs are localized on the
islands resulting in an insulating behavior of the film. The other scenario 
\cite{fermion} proposes localization of single electrons as a result of
disorder and Coulomb interaction effects.

The technique of quench condensation, i.e. sequential evaporation on a
cryogenically cold substrate under UHV conditions, has proven to be a very
useful tool for studying the SIT\cite{strongin,granular bob,granular
goldman,granular rich}. Depending on the choice of substrate, this technique
enables one to probe both mechanisms of the SIT: If the samples are quench
condensed on a passivated substrate (such as SiO), they grow in a granular manner so that
the film brakes up into separated islands. The average distance between the
islands decreases upon adding material. In these samples there is a critical
thickness $d_{c}$, below which no conductivity can be measured. For granular
Pb, $d_{c}\approx 100$\AA . Once the thickness, d, of the sample is larger
than $d_{c},$ the sheet resistance, $R$, drops exponentially with
thickness until, for $R\leq 6k\Omega ,$\ it switches to a
normal Ohmic behavior ($R\propto 1/d$). STM measurements on
quench condensed Pb granular films close to $d_{c}$ \cite{tony,phil
mag,valles stm} show that the grains are about 200\AA\ in diameter and
50-80\AA\ in height. On the other hand when the substrate is pre-coated by a
thin layer of amorphous Ge (which is insulating at 4K), the sample grows
uniformly and is continuous at a thickness of 1-2 monolayers of material\cite
{strongin} after which the sample obeys the usual $R\propto
1/d $ dependence. It is believed that the large number of dangling bonds in
the Ge layer act as nucleation centers for the evaporated metals, thus the
deposited adatoms are less mobile and the sample grows homogeneously.

These two geometries lead to two different types of SIT. Figure 1 compares
the behavior of uniform and granular quench condensed films of Pb on a SiO
substrate. 
\begin{figure}[b]
\centerline{
\epsfxsize=90mm
\epsfbox{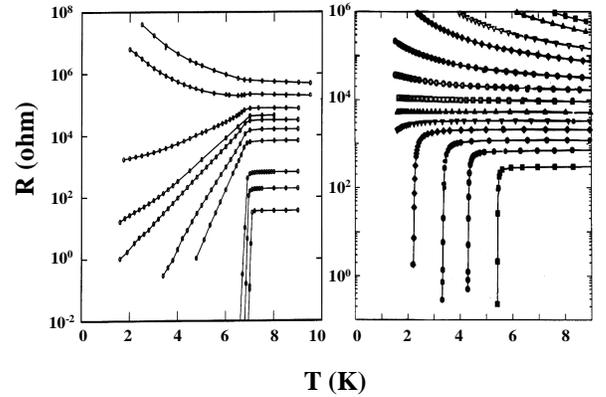}
}
\vspace{0mm}
\narrowtext
\caption{Resistance versus temperature for sequential layers of
quench-condensed granular Pb (left) and uniform Pb evaporated on a thin Ge
layer (right). Different curves correspond to different nominal thickness.
 }
\end{figure}

In the uniform film it is seen that \ for $R>6k\Omega $
the sample is insulating. As the film thickness is increased and the
resistance is decreased, a superconducting transition evolves with
increasing critical temperature, $T_{c},$ which approaches the bulk value
(7.2K for Pb) for thick enough samples. The superconducting transitions
themselves are relatively sharp and well defined. In the granular film a
crossover from an insulating to superconducting behavior is also observed.
However, here $T_{c}$ is not very well defined and the superconducting
transitions have broad tails which become sharper as material is added to
the film. If $T_{c}$ is defined as the temperature at which the resistance
starts dropping exponentially, it has bulk value and barely changes
throughout the entire SIT. Moreover, even on the insulating side of the SIT
the curve changes its slope at $T=T_{c}$ reflecting the influence of
superconductivity even in the thinnest measurable samples. A similar effect
is seen in the properties of the energy gap, $\Delta $. Tunneling
measurements show \cite{compare,tunneling} that while in the uniform film $%
\Delta $ grows with increasing thickness so that the ratio $\frac{\Delta }{%
T_{c}}$ is roughly constant, in the granular case the bulk energy gap is
measured even on the insulating side of the transition and its does not
change with additional material.

These differences are attributed to a different nature of the SIT\cite
{compare}. In the granular case the grains are large enough to sustain
superconductivity with bulk properties. However, for the high resistance
samples, there are phase fluctuations between the grains leading to an
insulating R-T curve. As more material is added the Josephson coupling
increases and eventually superconducting percolation is achieved. The situation is
different for the uniform geometry where the film is believed to be
homogeneous on an atomic level. In this case the SIT is due to suppression
of the order parameter amplitude at ultra-low thickness due to an interplay
between coulomb interactions and disorder.

So far, all studies on quench condensed based SIT have been performed on
either extreme of very homogeneous layers or granular films with grains large
enough to have bulk superconductor characteristics. In this paper we
describe results obtained on films that bridge these two extremes. We grow
granular Pb films in which the grains are too small to sustain bulk
superconducting properties and investigate the nature of the SIT in these
samples. This geometry is intriguing. Though much attention has been
directed towards understanding the nature of superconductivity in grains in
which the discrete electronic level spacing is larger than $\Delta ,\cite
{d>D,von delft}$, not much is known about an array of coupled ultrasmall
grains.

Since the Ge under-layer appears to be the cause for the uniformity of the
films, the natural way to try and reduce the size of the grains would be to
grow a very thin layer of Ge prior to the Pb evaporation. This turns out to
be very difficult because 1-2 monolayers of Ge are enough to cause
atomically uniform films. Alternatively, one can achieve a suitable
substrate for ultrasmall grains by applying different evaporation conditions
than those used in previous works. Conventional quench condensed films, such
as those described in figure 1, were deposited in a vacuum can set-up
immersed in a liquid He$^{4}$ bath, resulting in UHV conditions ($%
P<10^{-11}mbar$) due to cryopumping of the can. The samples described in
this work were grown in a normal high-vacuum evaporation chamber equipped
with a cold finger capable of reaching temperatures as low as 4K. The base
pressure in the chamber was $10^{-8}$ mbar. This pressure is low enough to allow
relatively pure evaporations but it is still a few orders of magnitude
higher then that in the previous set-ups. In this system one can expect
particles from the residual pressure in the chamber to cryopump on the
substrate, which is the coldest element in the set-up. As time goes by
nucleation centers form on the substrate, causing the granularity of the
evaporated material to decrease. Indeed, depending on the waiting time
before evaporation, we achieve samples with $d_{c}$ between 20 and $50$\AA .
This should be compared to $d_{c}\approx 100$\AA\ for granular Pb and $%
d_{c}\approx 5$\AA\ for uniform Pb.

The resistance versus temperature curves for several samples having $d_{c}\ $
between $45$ and $22$\AA\ are shown in figure 2. These curves were measured
using standard 4 probe AC techniques and making sure, for each point on the
curve, that the I-V characteristics are in the linear regime. An SIT is
observed for all these samples. However, the details of the transition are
different than both those of uniform and those of granular systems. In fact
the samples have characteristics of both types of geometries. On one hand $
T_{c}$ changes considerably as more material is added and on the other hand
the transitions exhibit broad tails.
\begin{figure}[b]
\centerline{
\epsfxsize=90mm
\epsfbox{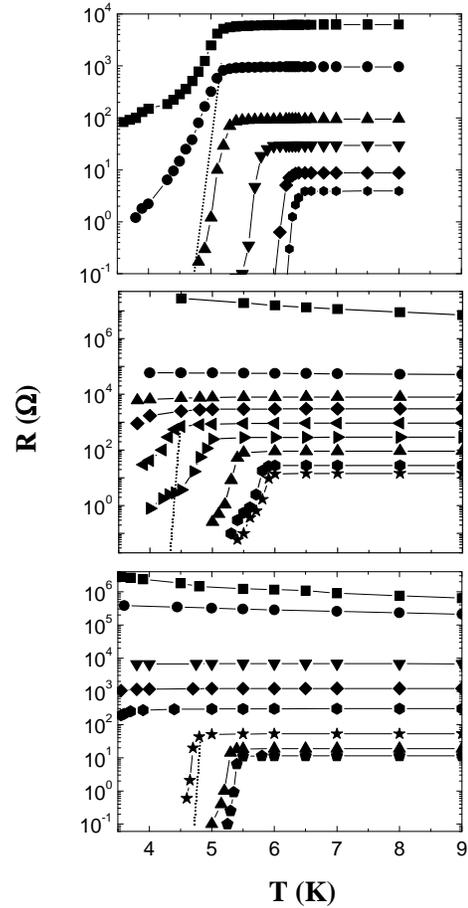}
}
\vspace{0mm}
\narrowtext
\caption{Resistance versus temperature of Pb quench condensed samples
having $d_{c}=45,$ 33 and 22\AA\ from top to bottom respectively. The
dashed lines are the calculated curves that are expected according to
equation 2 for large grains.
 }
\end{figure}

The fact that these samples are granular in nature can be seen from a number
of experimental findings. First, $d_{c}$ is much larger then a monolayer
indicating that the films are not atomically uniform. Secondly, the
resistance as a function of thickness (for $d>d_{c}$), as shown in figure 3,
begins with a sharp exponential dependence before turning Ohmic,
reflecting the tunneling nature of the transport. In addition the I-V curves
shown in the inset of figure 3, exhibit a series of hysteretic jumps,
typical to a granular system which contains an array of Josephson junctions
with different critical currents. These cause avalanche bursts of the I-V
curve at discrete voltages\cite{aviad1}.

\begin{figure}[b]
\centerline{
\epsfxsize=90mm
\epsfbox{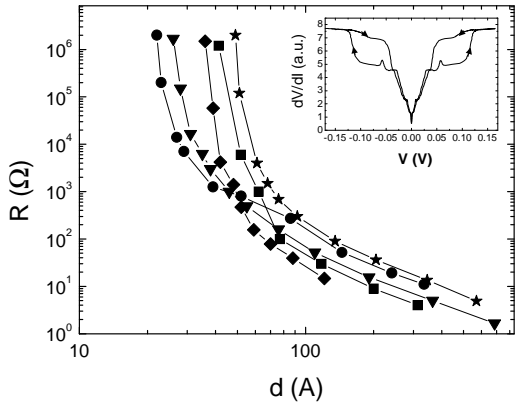}
}
\vspace{0mm}
\narrowtext
\caption{Resistance versus thickness for a number of samples having $
d_{c}$ of 22\AA\ (circles), 26\AA\ (triangles), 35\AA\
(diamonds), 44\AA\ (squares) and 49\AA\ (stars). Insert: Dynamic resistance
as a function of voltage for a typical sample measured at T=4K. The
measurements are taken sweeping the bias in both directions showing the
hysteretic nature of the curves. The arrows mark the direction of the sweeps.}
\end{figure}

Hence, the film is composed of grains. Nevertheless, the critical
temperatures we measure for all the layers are smaller than that of the
bulk. This together with the relatively small $d_{c}$ of the films lead to
the conclusion that the grains are considerably smaller then those in the
previously studied granular structures. Using a simple model based on the
geometries deferred from the STM measurement in the large grains we estimate
the average grain size in our samples to be around 40-80\AA\ in diameter and
about $50$\AA\ in height. The electronic level spacing, d$^{\ast }$ in a
single grain can be calculated using the free electron expression for the
density of states, $N(E_{F)}:$%
\begin{equation}
d^{\ast }=\frac{1}{N(E_{F})}=\frac{2\pi ^{2}\hbar ^{2}}{mk_{F}\Omega }
\label{level-spacing}
\end{equation}
where $\Omega $ is the volume of the grain. For our Pb grains this yields a
level spacing of $\sim $0.75 meV which is smaller than the superconducting
energy gap, $\Delta =1.4meV$. Thus, superconductivity can be expected to be
suppressed and the critical temperature of the grains may indeed be smaller
than that of the bulk.

Because the geometry of these systems lies between that of the
uniform films and the large-grain films, one could expect that the physical
properties would also show intermediate behavior. This is not the case.
Since d$_{c}$ is geometry and material dependent, the relevant parameter for
comparing different systems is the normal-state sheet resistance, R$_{N},$
which reflects the degree of disorder, rather than the film thickness. In
the large grain systems, the resistance versus temperature for $%
R_{N}<80k\Omega $ and $T<T_{c}$ follows an exponential behavior (see figure
1) and can be expressed in the following way \cite{lynne}:

\begin{equation}
R=R_{0}e^{\frac{T}{T_{0}}}  \label{R-T}
\end{equation}

Though the cause for this dependence is not yet understood, it turns out to
be universal for all quench condensed granular superconductors. Moreover,
the slope of the curves, $\frac{1}{T_{o}},$ depends only on $R_{N}$ and not
on the material or the $T_{c}$ \cite{ofer}. Our samples do not follow this
dependence. Instead, $R$ versus T is always slower than an
exponential law and the slope of the curves everywhere is less steep then
what it would be for a similar $R_{N}$ in the usual granular system. This
can be seen in figure 2 where we have sketched curves as they would have
been measured for large-grain systems having a similar $T_{c}$ and $R_{N}$. Clearly the
small-grain samples exhibit a much slower dependence on temperature. This
trend becomes weaker as $d_{c}$ decreases, as expected when the sample
approaches uniformity.

The critical temperatures measured in these samples also do not fall between
those of the two extreme geometries. Figure 4 depicts $T_{c}$ as a function
of $R_{N}$ for a number of our samples compared to that of uniform films 
\cite{compare}. It is seen that all samples with $R_{N}<1k\Omega $ have
critical temperatures smaller than that of the uniform film. The smaller the 
$d_{c}$ the smaller are the critical temperatures (though this trend seems
to stop as the sample approaches uniformity).

\begin{figure}[b]
\centerline{
\epsfxsize=90mm
\epsfbox{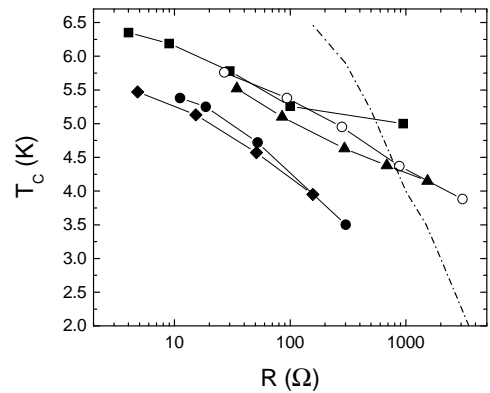}
}
\vspace{0mm}
\narrowtext
\caption{Critical temperature, defined as the temperature at which the
resistance drops to half its normal value, as a function of normal-state
sheet resistance for samples having $d_{c}$ of 22\AA\ (full circles), 26\AA\
(diamonds), 30\AA\ (triangles), 37\AA\ (empty circles) and 45\AA\ (squares).
The dashed line is the dependence for uniform films.
 }
\end{figure}    Hence, regarding the tail
broadness, the small grain films are more extreme than the large grain
films, while, concerning the critical temperatures they fall shorter than
those of uniform films.

The results presented above imply that the mechanism of the SIT in
the ultrasmall grain system is different than both the transition in a
uniform film, which is dominated by order parameter amplitude suppression,
and that of the large-grain systems which is governed by phase fluctuations.
Though the mechanism is not yet understood it is clear that fluctuations in
either the phase or the amplitude of the order parameter alone is not enough
to account for the findings. Rather, it seems reasonable that a combination
of both parameters are involved in the behavior of these systems. In order
to achieve superconductivity, two conditions have to be fulfilled. One is
that all the grains in the conduction percolation network would be close
enough to allow Josephson coupling and the second is that they all be
superconducting. Unlike the other two geometries where all the film turns
superconducting at once, the ultrasmall grain systems may involve a
distribution in $T_{c}$ due to a distribution in grain size. Hence, at
temperatures smaller than the bulk $T_{c}$ many of the grains are
non-superconducting while others may have turned superconducting. This leads
to a situation (which does not occur in the large-grain films) in which
grains can be close enough to allow strong Josephson coupling, and yet the
sample will not have long-range superconductivity because many of the grains
are not superconducting. Adding small amounts of material to the layer
causes the average grain size to grow (either by coalescence or by
increasing the tunneling between grains) thus increasing the average
critical temperature. Similarly, as the temperature of a layer is decreased
the portion of grains in which $T<T_{c}$ grows, increasing the
superconducting area in the film. This may be the reason for the relatively
weak temperature dependence observed in these samples because overcoming the
phase fluctuations is not enough for ensuring superconductivity in the
system. In this geometry, the dominating factor which drives the transition
may be the section of the film which is superconducting.

Finally, we note that the quench condensation technique enables one to use simple
systems as prototypes to study the SIT in more complicated systems. The use
of different substrates for various morphologies leads to different types of
SIT which provide much insight into the physical mechanisms determining the
transition. The samples described in this work demonstrate the nature of the
transition in systems in which very small regions of superconductivity are
embedded in a non-superconducting matrix. This is believed to be the case in
many composite samples studied recently (see for example \cite{gantmacher}).
The R-T dependences of any dirty superconductor can be used as a fingerprint
to determine the geometry of the superconducting material and the governing
SIT mechanism.

We gratefully acknowledge illuminating discussions with R. Berkovits, R.C.
Dynes, Z. Ovadyahu, M. Pollak and M. Schechter. We also thank VST company
for the design and manufacturing of the experimental system. This research
was supported by the Israeli academy of science.

\end{multicols}
\end{document}